\documentstyle[12pt,preprint,aps,tighten,epsfig,floats]{revtex}

\def\beq{\begin{equation}}
\def\eeq{\end{equation}}
\def\beqa{\begin{eqnarray}}
\def\eeqa{\end{eqnarray}}
\def\be{\begin{equation}}
\def\ee{\end{equation}}
\def\bea{\begin{eqnarray}}
\def\eea{\end{eqnarray}}
\begin{document}

\title{Random values of the cosmological constant}
\author{John F. Donoghue} 
\address{Department of Physics, University of Massachusetts\\
Amherst, Massachusetts 01003  \\ and \\ TH Division, CERN, Geneva}
\maketitle
\thispagestyle{empty}
\setcounter{page}{0}
\begin{abstract}

One way that an anthropic selection mechanism may be manifest in
a physical theory involves multiple domains in the universe with
different values of the physical parameters.
If this mechanism is to be relevant for
understanding the small observed value of the cosmological constant, 
it may involve a mechanism by which some contributions to the
cosmological constant can be fixed at a continuous range of values in 
the different
domains. I study the properties of
four possible mechanisms, including the possibility of the Hubble
damping of a scalar field with an extremely flat potential. 
Another interesting possibility involves
fixed random values of non-dynamical form fields, and a cosmological mechanism is 
suggested. This case raises the possibility of anthropic selection
of other parameters in addition. Further requirements needed for 
a consistent cosmology are discussed. 
\end{abstract}

\vspace{1.0in}

\section{Introduction}
  
  The problem of understanding a small but non-zero cosmological constant
  ($\Lambda$)\footnote{ If the cosmological constant is indeed the explanation
  of the recent supernova observations\cite{supernova}, 
the value would be $\Lambda = (1.2 \pm
  0.4) \times 10^{-123} M_P^4$, where $M_P = 1.22 \times 10^{19}$~GeV
is the Planck mass.} 
  appears even harder than it would be if the
  cosmological constant were identically zero\cite{cosmological}. 
There are many contributions to
  $\Lambda$, ranging from zero-point energies to Higgs and QCD vacuum
  condensates. Observing a non-zero value tells us that we should not seek a
  principle that requires these contributions to cancel exactly.
  However, empirically a
  partial cancellation must occur 
  and must be extremely fine-tuned in order to result in such
  a tiny residual.
  
The problem is so severe that it forces us to take seriously 
the anthropic\cite{anthropic}
multiple domain solution which would naturally lead to the observation of
a small non-zero 
$\Lambda$. Under this hypothesis\cite{weinberg1,linde0,lindebook,vilenkin0}, 
the cosmological constant is a parameter that
can take on different values in different domains of the universe, with an
assumed cosmological evolution such that we live entirely within a single
domain. Domains with ``normal'' values of the cosmological constant would
collapse quickly or expand exponentially rapidly and could not lead to life of
any form. Only those with a small enough
residual $\Lambda$ have the conditions appropriate for
life and it is only this restricted range that we should consider. Under this
hypothesis, we would expect to observe a non-zero value of $\Lambda$, 
since there is no
mechanism forcing it to be zero, and the magnitude would be expected 
to be typical of
the anthropically allowed range. Weinberg\cite{weinberg1,weinberg2} has
phrased this constraint in a physical way by asking about the mean value of
$\Lambda$ in universes in which matter clumps into galaxies, the clumping being
a needed precursor to life. Under plausible estimates\cite{estimates}, 
the observed value of
$\Lambda$ is reasonably typical of the mean viable value. 
Rees and Tegmark\cite{rees} have
pointed out that in a more general context there is an allowed two dimensional
area in the values of $Q$ and $\Lambda$, where $Q$ is the magnitude of the
initial density perturbations, again such
that our values are reasonably typical. While this hypothesis could provide a
natural explanation of the value of $\Lambda$, its physical foundation remains
unclear and we need to look for possible physical realizations. 

 For the mechanism to be contained in a physical theory there must be two main
 ingredients. The first is the generation of an appropriately large
 universe with domains that are presently disconnected. 
This is a relatively simple requirement.
 There are many available ideas for having 
 quantum fluctuations or random dynamics
 influence physics within one causally connected region of the early universe.
 Inflation\cite{lindebook,inflation}, or pre-big-bang evolution\cite{pbb} 
can then insure that we live entirely
 within a region which evolved from a 
 single such domain. Disconnected regions of
 the universe are a common occurrence in modern theories of cosmology.
 
 The difficult aspect of this hypothesis is
 contained in the second ingredient -
 the variability of physical parameters such as the cosmological constant. 
 Ordinarily, coupling constants and masses are constant parameters uniquely
 defined within a theory. However in this hypothesis, the
 requirement is that these parameters
 can take on multiple values, yet are essentially
 constant throughout our domain. The values of the parameters are related to
 the ground state of the theory. Different ground states correspond to
 differences of at least some parameters of the low energy theory.
 The usual situation envisioned in fundamental
 theories is that there is a unique ground state to the theory, or at most a
 discrete few ground states. Even in string theory, which classically has
 continuous families of ground states, one normally 
expects that non-perturbative effects
 will select at most a few possible true ground states per compactification. However,
it is unlikely that a set of discrete ground states is sufficient for implementing
the anthropic selection (see the next section). If we turn to continuously 
variable states there are difficulties in maintaining a stable set of parameters.
This paper examines issues associated with known mechanisms for 
implementing the multiple-domain/anthropic scenario. 

\section{Discrete versus continuous?}

The ground state of a theory like QCD appears to be unique. The electroweak 
theory has a continuous family of ground states, corresponding to different
directions of the Higgs field, but they are all equivalent and all have 
the same parameters. In more complicated theories with multiple Higgs fields
there can be several minima to the Higgs potential. These multiple minima
are potentially applicable to the multiple domain problem. 
One possibility is that some symmetry leads to the condition
where several ground states have the same energy. However, if they have the
same ground state energy then they have the same cosmological constant, and are
therefore not useful in this context. In the more common case where 
the minima are all of different energy, only one will be the true ground state. 
However, in some situations the time to tunnel from one ground state to the
true vacua can be long enough that the metastable states can be considered
in cosmology. These different ground states would correspond to 
different cosmological constants. 
Therefore it is reasonable to consider multiple metastable discrete 
ground states as a candidate mechanism in the multiple domain problem.

However, a few discrete ground states 
are not enough for the anthropic solution for the
 cosmological constant. A theory with multiple ground states must occur
at energies higher that that of the Standard Model. Let us denote the
scale of this future theory by $M_*$, with $M_* \ge 1$~TeV. The ground states
would be categorized by energies of this scale. In particular, the splitting
between the ground state with the smallest negative cosmological constant
and the smallest positive one would be of this size. It is extremely unlikely 
that a ground state would fall in the very tiny window that allows
a anthropically viable cosmological constant. That window corresponds to 
a range 
\beq
{(\Delta \Lambda)_{\rm anthropic} \over (\Delta \Lambda)_{\rm natural}}
 \sim 10^{-58} \left[ {1 ~TeV^4 \over M_*^4}\right]
\eeq
If a theory had a very densely packed set of states around $\Lambda =0$, 
it would contain an unnaturally small parameter describing the spacing
of these states (as well as possibly having difficulty arranging for these
states to be metastable for long periods).  The great disparity between
$M_*^4$ and $\Lambda$ indicates that one would require an additional 
mechanism to generate the possibility of fine tuning an anthropically
acceptable value\footnote{In fact, 
new mechanisms that may allow closely
spaced values for the cosmological constant have recently been
addressed in \cite{bp,wilczek2}}.

 The alternative is that the parameters can vary {\em continuously}, yet stay
 frozen at an arbitrary fixed value throughout our domain. 
Here the requirement is only that both signs of the cosmological constant be possible.
In this case,
 random dynamics
 will occasionally generate an acceptable $\Lambda$ in the neighborhood of
zero. However, this option is not
 without problems. 
 
 Let us consider the basic difficulty in a general framework.
 If the parameters can have continuously
 different values, they would be different in causally
 disconnected domains in the early universe. Therefore they can be 
 described by 
 space-time dependent fields. This means that we will always be looking at the
 dynamics of some fields. Since by assumption these fields are not
 constrained to be near a unique value, their potential, if they have any, must
 be small and they would normally be described as nearly-massless fields. 
 While inflation can readily lead to these fields
 becoming uniform throughout the observed universe, the difficult part is to
 understand why the dynamics of such a 
 field did not lead it to evolve towards a
 unique ground state. Therefore we are lead to consider fields whose dynamics
 have been frozen at continuous values 
in some fashion. This is the topic of the rest of this paper.

\section{Hubble damping}
   
There exists a simple mechanism that demonstrates that the freezing of dynamical
fields at random values is indeed possible. It is related to the ``slow roll''
mechanism which is important for inflation\footnote{However, 
it should be clearly stated
that the discussion which follows  does not apply to the field
responsible for inflation, but to a different scalar field that is being invoked
to address the cosmological constant problem.}. Consider a scalar field in an
expanding FRW universe governed by a scale factor $a(t)$. The equation of motion
for this field is 
\beq
\ddot{\phi} + 3H\dot{\phi} -{1\over a^2} \nabla^2 \phi = - V'(\phi) .
\eeq
where a dot denoted a derivative with respect to time and the prime denotes
differentiation with respect to the field $\phi$.  
$V(\phi)$ is the potential for the scalar field
and the Hubble parameter is defined by
\beq
H = {\dot{a(t)} \over a(t)}  .
\eeq
For a field which is sufficiently spatially uniform, one can drop the term
involving spatial gradients. In this case, a sufficiently flat potential, when
compared to the Hubble constant, will lead to $\dot{\phi} \sim 0$. Thus the
Hubble expansion can damp the time evolution of a uniform scalar field with a
sufficiently flat potential. The application of this mechanism to the
question of the anthropic solution to the cosmological constant problem
has been studied in detail in recent work by Garriga and Vilenken\cite{gv}, and 
discussed by Weinberg\cite{weinberg2}.

Let us look in more detail at the condition for the freezing of the field as we
will see that there
is a conflict related to the two uses of a flat potential in this scenario.
On one hand the potential must be flat with respect the {\em present} Hubble
parameter, which is a very small number, in order that the field be presently
frozen. This corresponds to the intuitive expectation that the Hubble expansion
plays very little role on the fields that we see around us, such that it must be
a {\em very} weakly varying potential if the Hubble parameter is 
to provide the damping to keep
the field from dynamically evolving. 
However, on the other hand, we need the potential to have enough variation that
its contribution to the vacuum energy is sufficient to influence the
cosmological constant. If the potential is too flat it contributes
too weakly to the cosmological constant to be able to nearly cancel the
other contributions to $\Lambda$.  
These dual requirements force certain unnatural
conditions on the potential and also pose an important requirement on the nature of
inflation.

The condition that the field remain
effectively frozen today means, 
among other things, that it is not changing so fast as
to contribute significantly to the present energy density. Thus its kinetic
energy is bounded\cite{gv}
\beq
 {1\over 2}{\dot{\phi}}^2  << \rho_0
\eeq 
where $\rho_0$ is the present energy density of the universe. This is related to
the present Hubble constant (neglecting a possible curvature contribution to the
Hubble constant) by
\beq
H_0^2 = {8 \pi \over 3 M_P^2} \rho_0
\eeq
Using the slow roll approximation, we find that this kinetic constraint implies
\beq
V'(\phi) \sim 3 H_0 \dot{\phi} << 3H_0 \sqrt{{3 M_P^2 H_0^2 \over 8\pi}}
\sim H_0^2 M_P \sim 10^{-122} M_P^3 . 
\eeq
Here we have use the kinetic energy bound, dropped constants of order unity and
used the value of the Hubble parameter $H_0 \sim 10^{-61} M_P$. 
The conclusion is that the potential must be {\em very} flat. 
However, this means that reasonable
variations of the magnitude of $\phi$ do not change the vacuum energy much.
Specifically, for this mechanism to be operable we need to be able
to nearly cancel the effect of other sources of vacuum energy,
which we can denote by $\Lambda_{\rm other}$. 
The variation of the vacuum energy
as we vary $\phi$ must then be of order $\Lambda_{\rm other}$. 
Let us distinguish two extreme situations: 
one where gravity or string theory 
provides the scale of the vacuum energy such that $\Lambda_{\rm other}\sim M_P^4$
and the other with low energy supersymmetry which 
implies $\Lambda_{\rm other}\sim 1$~TeV$^4$.
In terms of the potential, the requirement is
\beq
V'(\phi) \Delta\phi \sim \Lambda_{\rm other}
\eeq
Thus the very small values of $V'(\phi)$ can only be useful 
if $\phi \sim \Delta\phi$ is
very large. Inserting the constraint on $ V'(\phi)$ from Hubble damping
reveals just how large this value must be
\beq
\phi  >>  10^{122} M_P \left( {\Lambda_{\rm other} \over M_P^4} \right) 
\eeq
Even if we 
have
low energy supersymmetry
this leads to a strikingly large value of $\phi >> 10^{58} M_P$.

The extreme
flatness of the potential is a potential difficulty. 
Allowing quadratic and quartic
couplings, we will have
\beq
V'(\phi) = \mu^2 \phi + \lambda \phi^3 .
\eeq
Given the constraints on $\phi$ and $V'(\phi )$, we must have $\mu^2 < 10^{-244}
M_P^2$ and $\lambda < 10^{-488}$ in the case where 
$\Lambda_{\rm other}  $ is determined by the Planck scale
and $\mu^2 < 10^{-180}
M_P^2$ and $\lambda < 10^{-296}$ in the most favorable case of
weak scale supersymmetry breaking. At first sight this appears to be a fine
tuning which is even greater than that of the cosmological constant.
One might not
be worried about the flatness of the potential
since in supersymmetry flat potentials are ubiquitous, and one might
hope that this flatness could be preserved. When supersymmetry is
broken, radiative corrections will generate contributions to a potential.  
For a
potential this flat, all matter fields certainly need to be decoupled from the
$\phi$ field, or else they would generate a large potential after supersymmetry
breaking. The decoupling of matter fields is also required in order to
not violate general relativity constraints. The field $\phi$ is effectively
massless, given the flatness of the potential, and would lead to observable 
long range
forces if coupled to matter at even gravitational strength. However the
constraints from the lack of radiative corrections to the potential are even
stronger, and one is led to assume that matter fields can be completely
decoupled from $\phi$. This leads to the expectation that this field will not 
influence any of the other parameters of the Standard Model, as noted by 
Weinberg\cite{weinberg2}. It will be a formidable problem to generate a potential 
that is large enough to influence the cosmological constant, yet flat enough 
to not be presently evolving. It would be remarkable if the existence of
a viable domain is only possible due to the existence of such a extreme 
potential.

The needed initial conditions may also present a fine tuning
problem. 
The size of the field $\phi$ is not 
by itself is not the problem, since we have seen that
despite this large value the energy associated with the field is still below the
Planck mass. However for a field of this size to also 
have its {\em kinetic} energy below
the Planck scale requires an unnatural spatial and temporal constancy. In this
case the problem is not so much in 
the present epoch, when inflation could have smoothed out any
spatial variation in $\phi$, but in the early universe before the start of
inflation. At this time, in order that the kinetic energies not exceed the
Planck scale, we need the variation in time and in each of the spatial
directions to satisfy
\beq
\partial_0 \phi \sim \nabla_i \phi \sim {\phi \over L} < M_P^2
\label{density}
\eeq    
with $L$ being the scale factor which describes the constancy of the
field. (In an infinite universe $L$ would be the wavelength of the field
configuration.) If the scalar field evolves classically its
magnitude would be the same in the
early universe, and one is then constrained to have
\beqa
L & > & 10^{122} M_P^{-1} \sim 10^{80} ~{\rm light-years} \ 
\ \ \ \ ({\Lambda_{\rm other} \sim M_P})  \nonumber \\
& > & 10^{58} M_P^{-1} \sim 10^{16} ~{\rm light-years} \
\ \  \ \ ({\Lambda_{\rm other} \sim 1~{\rm TeV}^4})   
\eeqa
Thus the requirement is that
the field initially (at the start of inflation) have an extremely large value,
but have a incredibly tiny spatial and temporal variation. 
These initial values are quite unnatural, and tell us that the classical 
evolutions is not a natural solution.

However, quantum fluctuations during inflation can modify the field values, and 
if inflation is long enough, would remove the unnaturalness issue for the initial
conditions\cite{vilenkingarriga}. 
This then can be converted into a limit on the length of the inflationary
epoch. Quantum fluctuations behave differently in an exponentially expanding space time. 
Long wavelength modes can get redshifted 
such that they become almost flat, in which case 
the Hubble damping freezes them to constant values that add to the value 
of the classical field. Since the quantum fluctuations carry either sign,
this leads to a random walk character
for the net field values. Different regions in the inflating domain can 
thus develop different 
values of the field, with a rms deviation that grows as $\sqrt{t}$. The heuristic 
explanation is as follows, although the result is derived from more rigorous 
calculations\cite{desitter,eternal}. In each 
causally connected region of size $H^{-1}$, 
fluctuations are independent. (In this section $H$ refers to the Hubble constant
during the period of inflation, rather than in the present epoch.)  
In an expansion time of $H^{-1}$ a typical fluctuation is of 
size $\Delta\phi \sim {H/2\pi}$. Since the expansion can freeze this field,
over many expansion times these fluctuations
then add as a random walk, resulting in a spread of values of order
\beq
\delta \phi^2 = \left( {H \over 2\pi}\right)^2 Ht
\eeq
Because these fluctuations take place during inflation, and the expansion 
smooths out the spatial variation, the kinetic energy constraint is 
never violated. As long as the potential energy does not grow larger than 
$M_P$, any value of the field can be reached in some domain if inflation
goes on long enough.
Thus if the initial value of $\phi$ starts off as  $\phi \le M_P$ and
 $H,\Lambda_{\rm other}$ are also of order $M_P$, one requires $N=Ht \sim 10^{244}$ 
e-foldings of inflation in order to have quantum fluctuations allow the field to 
grow to sufficient size to be relevant for the anthropic constraint. For the case where 
all of these quantities are as small as 1 TeV, the required number of e-foldings is 
$10^{148}$. If a theory has only 60-100 e-foldings, the quantum fluctuations cannot
solve the initial value problem. 
However the constraint 
on the amount of inflation
can be solved naturally in the various versions of eternal inflation\cite{eternal},
in which inflating domains continue forever, and our domain can 
have undergone an unlimited 
amount of inflation. Therefore, the anthropic mechanism is most naturally embedded
in theories of eternal inflation, as in \cite{gv}.

The multiple domain 
hypothesis raises the possibility of naturally providing a way to solve the 
fine tuning problem. 
The Hubble damping mechanism is interesting because it demonstrates that
fields can become frozen at a continuous range of values.
The difficulty with the extremely flat potential
can be traced back to the the reliance on 
the Hubble term in the equation of motion to provide the mechanism for freezing
the field. At present, $H$ is too small to provide much 
influence on the behavior of fields. It is useful to 
search for more efficient methods of damping the dynamics.

\section{Kinetic freezing}

We could ask why, in the previous analysis, we could not simply redefine 
the scalar field by an overall scale such that its magnitude looks more normal,
at the expense of also redefining parameters in the potential. The answer is
that the condition that set the 
scale was the requirement that the kinetic terms
be conventionally normalized. This suggests that by playing with the overall 
factor in front of the kinetic energy, one could also freeze the dynamics.
In fact, this idea has been suggested in the context of hyperextended 
inflation\cite{steinhardt1}, and a variant has recently been invoked to control the
dilaton potential\cite{steinhardt2}.
In this situation, one imagines that nonrenormalizable interactions are 
present in
the action, such that the lagrangian becomes
\beq
{\cal L} = {1\over 2}f(\phi ,\psi) \partial_\mu \phi \partial^\mu \phi +
V(\phi, \psi) + \ldots
\eeq
The function $f(\phi,\psi)$ is a unknown function that can depend on $\phi$ and
on other fields, here labeled $\psi$.
In this case, at values where $f$ is large, the
fields are effectively frozen even if fields are not at the minimum of the
potential. 
This can be seen from the equation of motion for $\phi$
\beq
f (\ddot{\phi} + 3H\dot{\phi}) +{1\over 2} f'\dot{\phi}^2 = - V'(\phi) .
\eeq
where
\beq
f' = {\partial f(\phi,\psi)\over \partial \phi}
\eeq
If $f$ or $f'$ is 
large, this can be a mechanism for slowing
further dynamical evolution, yet it is problematic when applied to the cosmological
constant.

The goal here is to allow a a more natural size of the potential.
Using notation from the Sec. III this
means that a potential of size
\beq
V'(\phi) \sim {\Lambda_{\rm other} \over M_P}
\eeq
will allow $V'(\phi)\Delta\phi \sim \Lambda_{\rm other}$ for $\phi$ ranging over a
natural range $\Delta \phi \sim M_P$. (Here we will not worry about a few extra
powers of ten).
Since the smallest reasonable expectation for $\Lambda_{\rm other}$ is of order
the scale of low energy supersymmetry, in the absence of other mechanisms,
this means that we need a potential of rough size
\beq
V'(\phi) \sim {1 ~{\rm TeV}^4 \over M_P} \sim 10^{-64} M_P^3   \ \ .
\label{constraint}
\eeq
Combining the equation of motion with the constraint on $\dot{\phi}$,
this means that we need
\beq
f' > 10^{58} M_P^{-1}
\eeq
While this mechanism may also be used to freeze the fields
it is questionable whether 
it is reasonable to get non-renormalizable terms so large. 
In an effective field theory description, non renormalizable terms 
occur as small corrections to the basic theory, due to interactions
with degrees of freedom which are much heavier. The expectation of 
effective field theories, born out in known examples, is that once the 
nonrenomrmalizable terms are of order unity, we excite the high energy 
degrees of freedom directly and the theory changes to a new effective
theory in which these fields are dynamical variables.
It is not natural to achieve such extremely large nonrenormalizable
interactions.

In fact, one can see that this is related to the mechanism of  
the previous section
in the 
special situation where $f$ either does not depend on other fields $\psi$, or 
these fields are held fixed at the minimum of a potential, $\psi = <\psi>$,
and an integrability constraint is satisfied.
In this case a field redefinition changes the problem exactly back to the situation
of the previous section. Define
\beqa
\chi & = & g(\phi)   \nonumber \\
\partial_\mu \chi & = & g'(\phi) \partial_\mu \phi
\eeqa
If we then identify
\beq
 g'(\phi) = f^{1\over 2}(\phi, <\psi>)
\eeq
and this can be integrated to obtain $g$, the Lagrangian is transformed into
\beq
{\cal L} = {1\over 2} \partial_\mu \chi \partial^\mu \chi +
V(g^{-1}(\chi)) + \ldots
\eeq
This is just a conventionally normalized action with a suppressed potential.

\section{Radiative damping}

Finally, what about other forms of damping? It is also possible to
damp the motion of a field through the radiation of particles. In effect, 
a changing field can produce particles, which takes energy out of the
field and hence slows down the rate of change. This effect has been 
studied in Ref \cite{wilczek1} and is used in the theory of ``warm
inflation''\cite{berera} Let us consider the
equations of motion
\beq
\ddot{\phi} +(\Gamma +3H) \dot{\phi}  = -V'(\phi)
 \eeq
with some unspecified damping $\Gamma$. This structure arises from the coupling
of the field to other particles, with the radiation of the other particles  
damping
the dynamics of the field. The proportionality of the damping 
to $\dot{\phi}$  is indicative that if the field is
not changing it does not radiate. Perhaps this mechanism 
could lead to naturally frozen
fields.

The difficulty in this case comes from the fact that 
$\Gamma$ must be very small at present.  Can we have $\Gamma \dot{\phi}$ as
large as $10^{-64}M_P^3$ as required by the constraint of Eq. \ref{constraint}?
The constraint on $\Gamma$ comes from
the generation of particles in the universe. The equation of motion is
equivalent to the conservation of energy in a co-moving volume $a^3$
\beq
{d \over dt}( a^3(t) \rho) = -p {d \over dt} a^3 - 
{\Gamma \dot{\phi}^2  } 
\eeq  
with energy density and pressure
\beqa
\rho & = & {1 \over 2} \dot{\phi}^2 + V(\phi)  \nonumber  \\
p & = & {1 \over 2} \dot{\phi}^2 - V(\phi)
\eeqa
such that $\Gamma\dot{\phi}^2$ 
represents the rate of energy flow out of the field
$\phi$. Higher power dependence on $\dot{\phi}$ will not
change our argument.   
A reasonably loose constraint on the rate of energy production is that it is
smaller that the production of the full present energy in one Hubble time.
\beq
\Gamma \dot{\phi}^2 < H \rho_0 \sim 10^{-183} M_P^5
\eeq    
This then lets us put a constraint on the damping term in the equation of motion
using 
\beq
\Gamma \dot{\phi} = (\Gamma \dot{\phi}^2 \Gamma)^{1\over2} < 10^{-91} M_P^3 
({\Gamma \over M_P})^{1/2} .
\eeq     
Even if this unspecified damping mechanism was able to produce $\Gamma \sim
M_P$, this fails by 27 orders of magnitude to provide enough damping to
allow a reasonably sized potential.
        
\section{Form fields}

We may also turn to other ideas for fields
with frozen dynamics. Another possibility is known in the supergravity
literature, as first pointed out by \cite{form1,form2}
 Consider a field like a gauge potential but with three totally antisymmetric
Lorentz indices
\beq
A_{\alpha\beta\gamma} (x) = - A_{\beta\alpha\gamma} = -A_{\gamma\beta\alpha}
\eeq
such that its field strength tensor is also formed antisymmetrically
\beq
F_{\alpha\beta\gamma\delta} = \partial_{[\alpha} A_{\beta\gamma\delta ]}
\eeq
where the square brackets denote the antisymmetrization of the indices.
The Bianchi identity
\beq
 \partial_{[\alpha} F_{\beta\gamma\delta\rho]} = 0
\eeq
is then always satisfied in 4 dimensions since there is no totally antisymmetric
object with five Lorentz indices. The action
\beq
S_F ={-1 \over 48} \int d^4x {\sqrt{-g}} 
~F_{\alpha\beta\gamma\delta}F^{\alpha\beta\gamma\delta}
\eeq
leads to the equation of motion
\beq
\partial^\alpha \left[ \sqrt{-g} F_{\alpha\beta\gamma\delta}\right] = 0    .
\eeq
The only solution to this is 
\beq
F_{\alpha\beta\gamma\delta} = 
{c \over \sqrt{-g}} \epsilon_{\alpha\beta\gamma\delta}
\eeq
for arbitrary c. Thus this field is nondynamical, with only a constant solution.
Substitution of this solution in Einstein's equations
 shows that it behaves as a
positive cosmological constant. In the
language of differential forms, $A$ is a 3-form potential, and $F$ a 4-form
field strength, with equations of motion and Bianchi identity
\beqa
d * F_4 & = &0 \\
d F_4 & = & 0   .
\eeqa 
Form fields appear in the low energy limit of string theory and
M theory. The most obvious is the type II supergravity in the low energy limit
of M theory, where the 4-form field strengths occur explicitly. However, they
can also be obtained by dimensional reduction from higher form fields. Consider
a form field strength with more than four indices,
$F_{\alpha\beta\gamma\delta ....\rho}$. Upon compactification, some of the
indices can be assigned to the compact directions, becoming internal indices.
The number of such 4-forms will depend on the
particular number and symmetries of the compact subspaces.
Four-forms may also appear from lower dimension forms.
For an n-form in d dimensions, its dual is a d-n form.
Likewise duality
relates a 4- form in 4-d to a zero-form -i.e. a constant.

Hawking and Turok\cite{turok} 
have proposed the generation of a non-zero 4-form through
a tunneling mechanism involving an special instanton in the case of an open
universe. This calculation remains controversial, with a dispute over the
meaning of the instanton solution\cite{vilenkin}. The mechanism has
phenomenological problems, as it naturally predicts an almost empty universe.
Moreover, the mechanism generates one value of $\Lambda$ through out the entire
universe, such that the naturalness of
the anthropic selection is lost, and it does not correspond to the multiple
domain structure under consideration here.

If a multiple domain structure is to be realized in nature, it will 
be generated in the early universe. Therefore we should look to cosmology
for possible mechanisms. Here we suggest a mechanism which exploits the 
dimensional reduction that may take place in string theories 
The non-dynamical nature of the 4-form fields is only true in four dimensions.
In higher dimensions, the equations of motion allow the usual plane wave
solutions. The lack of dynamics in 4-d results from the restriction that the
Lorentz indices and the space-time 
variability lie entirely within the 4-d space.
This suggests a potential cosmological mechanism for the generation of the
4-form. Consider higher dimensional theories where compactification leads to a
4-d low energy theory. If cosmology goes through a phase where fields above the
compactification scale are excited at some time in the early universe, 4-form
fields will be dynamical. They will have fluctuating values, with a non-zero rms
field strength. As the universe expands and the average energy decreases, the
Kaluza Klein modes with excitations in the compact dimensions will decouple
leaving an effective four dimensional theory. As this transition occurs, the
4-form fields will become non-dynamical and will be frozen into random values in
different space-time regions. As the universe evolves to lower energies, these
values remain frozen. When
supplemented by inflation, such that we see only the field from a very small
initial patch, this can result in the multiple domain scenario.

In string theory there appears to be a barrier to the use of form
fields to generate random values of $\Lambda$. In a string theory ground state,
the values of the form field strengths are 
quantized\cite{bp,polchinski,polstrom}. This occurs because 
there are both electric and magnetic charges
coupled to the form fields. By analogy to the usual electric and magnetic
charges, these charges are quantized. Construction of various Gaussian surfaces
then imply that the flux, and hence the magnitudes of the constant form fields,
are also quantized. The cosmological mechanism described above could also
generate different values of the quantized form fields, but it might
appear that unless the 
size of the quanta 
are extremely small, the likelihood of solving the cosmological constant
problem is small\footnote{Of course it is also possible to imagine form fields without
invoking string theory. It has also been argued that there can be form fields
which are not coupled to string theory charges\cite{gabadadze}.}.

However the quantization constraint can still allow the form fields
to take on all values in a continuous range providing other fields adjust 
accordingly. 
The quantization constraint involves $V_7$, the volume of the 
compact seven dimensional manifold\cite{bp}. There are also additive 
contributions from possible flat background gauge potentials\cite{witten}
and constant fermion densities\footnote{This can be seen from the supergravity
equations of motion.}. 
If these were all to attain their low-energy values first, then 
the form field condensate would be forced to certain discrete
values. However, in the early universe the moduli controlling 
$V_7$, the gauge potentials
and the fermions are fluctuating.
The form field can take on any continuous value as
long as the other fields are adjusted to values consistent with the quantization
constraint.
As the universe cools to lower energies, the form field will become 
non-dynamical and will stay at its constant value. 
At low energies the potentials for the moduli and other
fields will become important and will approach their zero-temperature form.
These fields will then seek the minimum of their potentials, with
the quantization constraint being a constraint on what values are possible.
On other words, the form field value will become a constraint on vacuum
selection because it is no longer able to evolve. This inverts the 
usual reasoning, with the result 
that the form fields could end up at any value but the 
vacuum state adjusts in order to satisfy the quantization 
condition.

The frozen fields will have two effects. First, they can contribute directly to
the cosmological constant. However, there is also an indirect secondary effect
through the dilaton and moduli fields. As emphasized in Ref\cite{low}, 
the form
fields which carry string theory charges 
can influence the potentials for the moduli and dilaton fields. The
moduli and dilaton potentials vanish perturbatively, yet it is expected that
non-perturbative effects will generate potentials for these fields. The frozen
background of form fields will give additive contributions to the potentials.
This would amount to random shifts in the moduli potentials in different
domains, and would influence the ground state solution and the parameters
of the low energy theory. This
will then provide a further shift in the ultimate cosmological constant, since
every mass and coupling contributes to some extent to the vacuum energy. 
The influence on the moduli values may lead to the expectation
that other parameters in the theory also are variable.

In general, non-zero values of the form fields break supersymmetry. It is
known that there are special combinations of compactification and vacuum
expectation values that allow the existence of low-energy 
supersymmetry\cite{low,meissner}. Whether it is natural that such special
situations occur in the early universe is an open question. However, it may even
be preferable that the supersymmetry is broken at high scales (depending
ultimately on the
outcome of future experiments, of course.) In theories with random coupling
constants,
the fine tuning problem of the Higgs vev may not be the most serious issue. As
with the cosmological constant, there is a plausible anthropic constraint such
that we would only live in regions with a small Higgs vev\cite{agrawal}. This
occurs because if the vev is much larger than observed, the elements other 
than hydrogen do not exist and we lack the complexity needed for life. The
variability of the form fields and the moduli could allow the realization of
this anthropic constraint also.
Moreover, low energy supersymmetry poses significant problems for cosmology.
Scalar particles with TeV scale masses are ubiquitous in such theories, and the
dilaton in particular is model independent. These particles
dominate the energy density of the universe for so long that they spoils
nucleosynthesis\cite{ppp}. This problem, a string variant of the Polonyi
problem, has proven difficult to overcome. Moreover, it appears
difficult to implement inflation in theories with low energy
supergravity\cite{riotto}.  So cosmology may
welcome the situation where supersymmetry is broken at a high scale.

Let us then summarize the ingredients of a cosmology that would make use of this
mechanism. The first obvious requirement is that the evolution of the universe
must involve an early period where energies above the compactification scale are
excited. This is needed in order to excite the form fields. The required
features of compactification has not yet been studied much because most analyses
have been done under the assumption that supersymmetry survives to low energy.
So we don't yet know the full possibility for the field content below the scale
of non-supersymmetric compactifications. However, the supersymmetric
spectrum above the compactification scale has 
many fields, the dilaton and moduli, that have the
possibility of playing the role of the inflaton\cite{inflaton}. Use of these fields
would likely be possible if inflation and compactification occur at the same
scale. Finally we clearly need sufficient inflation to smooth out any initial
gradients in the fields.

\section{summary}
    This paper is a preliminary investigation into the field theory dynamics
that could lead to continuous 
random contributions to the cosmological constant in theories
with multiple domains with different parameters. Damping mechanisms appear 
to require rather extreme values for the potentials, the fields and/or the 
nonrenormalizable interactions. As noted by Weinberg\cite{weinberg2}, 
the need to decouple
all other fields from this scalar field, in order to preserve the flatness of 
the potential, has the consequence that it will not influence other parameters 
in the theory - that the cosmological constant will be the only parameter for
which an anthropic constraint is relevant

However
4-form fields appear as a quite natural 
 mechanism. For this to be applicable, we would want an energetic initial
 condition, to excite the form fields, and a inflationary phase to
 generate the uniformity of the observed universe.
 The fact that the form fields also influence the dilaton and moduli fields of
 string theory is also interesting. This would generate a chaotic component to
 the vacuum selection procedure and would thus influence the other parameters in
 the theory also. This may then also for the Higgs vev fine-tuning problem. 
There exists the possibility of testing the distribution of some of the
parameters through
the weight of the quark mass distribution\cite{weight}.  It remains to be seen
whether a fully complete model along these lines may be developed.

This paper has explored the situation in which the field variables influencing
the cosmological constant are
continuous. In this situation it is quite natural
that the cosmological constant
should occasionally be close enough to zero to satisfy Weinberg's 
anthropic constraint. 
In a recent paper, Bousso and Polchinski\cite{bp} 
have addressed the situation where 
multiple form fields can plausibly lead to discrete but closely spaced values for
the cosmological constant appropriate for an anthropic selection. 
The spacing of the values with separation of order $10^{-122}M_P^4$
appears to require very large internal dimensions or very many (of order 100)
form fields. 

\section*{Acknowledgments}
I would like to thank David Kastor, Renata Kallosh,
Andrei Linde, Joe Polchinski and especially
Daniel Waldram, Jaume Garriga and Alexander Vilenkin
for useful conversations. I also thank CERN for their kind
hospitality during most of time devoted to this work. This work 
has been supported in part by the U.S. National Science Foundation and by the
John Templeton Foundation.


\begin{thebibliography}{99}
\bibitem{vilenkingarriga} I would like to thank Jaume Garriga and Alex Vilenkin 
for pointing out the importance of quantum fluctuations 
and the relevance of eternal inflation
for the problem
of the size of the field $\phi$.
\end{thebibliography}
\end{document}